\documentclass[a4paper,reqno,11pt]{amsart}
\usepackage {amssymb}
\usepackage {amsmath,cite}
\usepackage {graphicx}
 \title {HEISENBERG ALGEBRA, UMBRAL CALCULUS AND ORTHOGONAL POLYNOMIALS}

\author{ G. Dattoli}
\address{ENEA, Dipartimento FIM, CRE Frascati.
C. P. 65, 000044 Frascati, Rome, Italy\\}
 \email{dattoli@frascati.enea.it}
 
 \author{ D. Levi}
\address{Dipartimento di Ingegneria Elettronica \\
Universit\'a degli
 Studi Roma Tre and Sezione INFN, Roma Tre\\
Via della Vasca Navale 84,
 00146 Roma, Italy\\}
 \email{levi@roma3.infn.it}
 
 \author{ P. Winternitz}
\address{Centre de recherches math\'ematiques and Department de math\'ematiques et de statistiques, Universit\'e de Montr\'eal, C.P. 6128--Centre Ville, Montr\'eal, QC H3C 3J7, Canada}
 \email{wintern@crm.umontreal.ca}
 \date{\today}

 \def\be{\begin{equation}}
\def\ee{\end{equation}}
 \def\ba{\begin{array}}
 \def\ea{\end{array}}
 \def\bea{\begin{eqnarray}}
 \def\eea{\end{eqnarray}}
 \def\beas{\begin{subequations}}
 \def\eeas{\end{subequations}}
 \def\bean{\begin{eqnarray*}}
 \def\eean{\end{eqnarray*}}

 \newtheorem{theorem}{Theorem}

\begin{document}
 \maketitle
\begin{abstract}
Umbral calculus can be viewed as an abstract theory of the Heisenberg commutation 
relation $[\hat P,\hat M]=1$. In ordinary quantum mechanics $\hat P$ is the derivative and $\hat M$ the coordinate 
operator. Here we shall realize $\hat P$ as a second order differential operator and $\hat M$ as a 
first order integral one.
We show that this makes it possible to solve large classes of differential and 
integro-differential equations and to introduce new classes of orthogonal polynomials, 
related to Laguerre polynomials. These polynomials are particularly well suited for 
describing so called flatenned beams in
laser theory

 \end{abstract}

 \section{Introduction}
The purpose of this article is to construct a realization of the Heisenberg 
algebra in terms of a second order differential operator $\hat P$ and a first order 
integral one $\hat M$.  We then use this realization to construct families of 
orthogonal polynomials and study their properties. Finally we discuss 
applications of these polynomials in mathematics (to solve 
integro-differential equations), optics (to describe flattened beams 
in optics) and in astrophysics (to describe distortions of the microwave background radiation).

This article is directly related to several research programs. The most 
general is that of umbral calculus, also known as finite operator calculus \cite{1,2,3,4}. This has been described in many 
different ways. Here we view umbral calculus as an abstract theory of the 
Heisenberg relation ${\left[ {\hat {P},\hat {M}} \right]} = \hat {1}$ and 
its mathematical implications.

A related direction is that of {\it monomiality}, the essence of which is a 
systematic investigation of the relations
\bea
\label{eq1}
 \hat {P}\,u_{n} = n\,u_{n - 1} , \qquad
 \hat {M}\,u_{n} = u_{n + 1}, 
 \eea
 leading to the representation of solutions of differential, difference and 
other equations as {\it monomials} $u_{n} = \,\hat {M}^{n}u_{0}$ \cite{5,6,7}.

The same algebra is of course obeyed by standard creation-annihilation 
operators in quantum physics. Monomiality and umbral calculus are closely 
related to the theory of Fock space ladder operators. For efficient recent 
applications of these concepts in mathematical physics see e. g. \cite{5,6,7,8,9,10,11}. 

A 
recent article \cite{12} was devoted to a systematic study of the realization of 
$\hat P$ as a difference operator and $\hat M$ as an operator the projection of which is the 
coordinate $x$. This was applied to show that continuous symmetries like Lorentz 
or Galilei invariance can be implemented in quantum theories on lattices 
\cite{12,13}.

In this article we systematically investigate a different realization, 
namely one in which $\hat P$ is a  second order linear differential operator 
in one real variable and $\hat M$ is a general first order linear integral operator.

The problem is formulated mathematically in Section 2 where we also obtain 
the general form of the operators $\hat P, \hat M$ in terms of one arbitrary function $Y(x)$ and 
four constants. We obtain the boundary condition for functions $u(x)$ defining the 
domain of the operators $\hat P$ and $\hat M$.

Section 3 is devoted to the basis functions $u_{n} = \,\hat {M}^{n}u_{0} $, 
where $u_{0}$ is the {\it seed} or {\it vacuum} function. 

We show that we can either choose $u_{0} = 1$ and then restrict the form of 
the function $Y(x)$ or leave $Y(x)$ arbitrary and choose $u_{0} (x) = {Y}'(x)\,(Y(x) + 
c_{0})^{\alpha}$ where $c_{0}$ and  $\alpha$ are  
constants. We show that the basis functions $u_{n} (x)$ are eigenfunctions of 
a second order linear differential operator $L $ with integer eigenvalues $\lambda 
_{n} = \,n + 1$. The operator $L$ is {\it factorized} in the sense that we have 
$L = \hat P \hat M$, where $\hat P$, $\hat M$ are the differential and integral operators  constructed 
in Section 2, respectively.

In Section 4 we show that the eigenfunctions $u_{n}$ form an orthogonal set 
and obtain the corresponding measure and integration limits.

The explicit form of the eigenfunctions is obtained in Section 5 for two 
different cases, namely $u_{0} = 1$ and $u_{0} (x) = {Y}'(x)\,Y(x)^{\alpha 
}$, where $\alpha$ is a constant. In both cases the eigenfunctions are 
expressed in terms of Laguerre functions of non-trivial arguments.

Section 6 is devoted to three types of applications: the solution of 
integro-differential equations,  the description of flattened beams in 
optics and the distortion of background radiation in astrophysics.

Conclusions and future outlook are presented in the final Section 7.

\section{LINEAR DIFFERENTIAL AND INTEGRAL OPERATORS SATISFYING THE 
HEISENBERG ALGEBRA}

Let us consider two linear operators that we postulate to have the following 
form
\begin{equation}
\label{eq2}
\begin{array}{l}
 \hat {P} = \Phi _{2} (x)\,\hat {D}_{x}^{2} + \Phi _{1} (x)\,\hat {D}_{x} + 
\Phi _{0} (x), \\ 
 \hat {M} = g(x)\,\hat {D}_{x}^{ - 1} + k(x) \\ 
 \end{array}
\end{equation}
where
\begin{equation}
\label{eq3}
\begin{array}{l}
 \hat {D}_{x} \equiv {\frac{{d}}{{d\,x}}}, \\ 
 \hat {D}_{x}^{ - 1} u(x) \equiv {\int\limits_{x_{0}} ^{x} {u(x)\,dx}}  \\ 
 \end{array}
\end{equation}
and $\Phi _{2,\,1,\,0} (x)$, $g(x)$, $k(x)$ are some sufficiently smooth 
functions, to be determined below.

We impose the following requirements
\begin{enumerate}
\item The products $\hat {P}\,\hat {M}$ and $\hat {M}\,\hat {P}$ are differential operators
\item Their commutator satisfies the Heisenberg relation
\begin{equation}
\label{eq4}
{\left[ {\hat {P},\hat {M}} \right]} = \hat {1}.
\end{equation}
\end{enumerate}
The product $\hat {P}\,\hat {M}$ will a priori contain a term proportional 
to $\hat {D}_{x}^{ - 1}$. The condition for it to be absent is 
\begin{equation}
\label{eq5}
\Phi _{2} (x)\,{g}''(x) + \Phi _{1} (x)\,{g}'(x) + \Phi _{0} (x)\,g(x) = 0,
\end{equation}
where the primes denote derivatives with respect to the argument.

The product $\,\hat {M}\,\hat {P}$ and hence also the commutator ${\left[ 
{\hat {P},\,\hat {M}} \right]}$ will apriori contain anti-derivatives $\hat 
{D}_{x}^{ - k} $of all orders. Indeed, the use of the Leibnitz formula 
extended to anti-derivatives yields
\begin{equation}
\label{eq6}
\hat {D}_{x}^{ - 1} {\left[ {f(x)\,u(x)} \right]} = {\sum\limits_{r = 
0}^{\infty}  {( - 1)^{r}}} (D^r f(x))\,D^{-(1+r)}u(x),
\end{equation}
where $f(x)$ can be a function or an $ x$--dependent operator.
Calculating the product $\hat {M}\,\hat {P}$we obtain
\begin{equation}
\label{eq8}
\begin{array}{l}
 \hat {M}\,\hat {P}\,u = k(x)\,[ {\Phi _{2} (x)\,\hat {D}_{x}^{2} + 
\Phi _{1} (x)\,\hat {D}_{x} + \Phi _{0} (x)} ]\,u(x)\, + 
\,g(x)\,[ \Phi _{2} (x)\,\hat {D}_{x} - \\ {\Phi} '_{2} (x)\,\, + \,\Phi 
_{1} (x) ]\,\,  
 + \,[{\Phi} ''_{2} (x)\, - \,{\Phi} '_{1} (x)\, + \,\Phi _{0} (x)]\,\hat 
{D}_{x}^{ - 1} + ...\vert _{x_{0}} ^{x} \\ 
 \end{array}
\end{equation}
which will be a differential operators if
\begin{equation}
\label{eq9}
{\Phi} ''_{2} (x)\, - \,{\Phi} '_{1} (x)\, + \,\Phi _{0} (x) = 0.
\end{equation}
The  coefficients of $D^{-k}$ with $k \geqslant 2$ are all equal to $0$
(as differential consequences of eq. (\ref{eq9})).

The commutation relation (\ref{eq4}) then imposes two further equations
\begin{equation}
\label{eq10}
\begin{array}{l}
 2\,\Phi _{2} (x)\,{g}'(x)\, + \,{\Phi} '_{2} (x)\,g(x) = 1 \\ 
 {k}'(x) = 0, \\ 
 \end{array}
\end{equation}
and a boundary condition at $x = x_{0} $, namely
\begin{equation}
\label{eq11}
[\Phi _{2} (x)\,{u}'(x)\, - \,{\Phi} '_{2} (x)\,u(x)\, + \,\Phi _{1} 
(x)\,u(x)] \vert _{x = x_{0}}  = 0.
\end{equation}
Solving the first of eqs. (\ref{eq10}) either for $\Phi _{2} (x)$ or $g(x)$, we 
obtain
\begin{equation}
\label{eq12}
\begin{array}{l}
 g(x) = {\frac{{1}}{{2\,\sqrt {\Phi _{2} (x)}} }}{\left[ 
{{\int\limits_{x_{0}} ^{x} {{\frac{{d\xi} }{{\sqrt {\Phi _{2} (\xi )}} }} + 
g_{0}} } } \right]}, \\ 
\mbox{or} \\
 \Phi _{2} (x) = {\frac{{1}}{{{\left[ {g(x)} \right]}\,^{2}}}}{\left[ 
{{\int\limits_{x_{0}} ^{x} {g(\eta )\,d\,\eta + c_{0}} } } \right]}, \\ 
 g_{0} = 2\,\sqrt {c_{0}} , \\ 
 \end{array}
\end{equation}
where $c_{0}$ is an integration constant. Furthermore from the 
integration of the second of eqs. (\ref{eq10}) we find 
\begin{equation}
\label{eq13}
k(x) = k_{0} 
\end{equation}
with $k_{0} $ a constant. Eqs. (\ref{eq5}) and (\ref{eq9}) provide expressions for 
$\Phi _{1} (x)$ and $\Phi _{0} (x)$ in terms of $\Phi _{2} (x)$ and $g(x)$.

We shall use the second of eqs. (\ref{eq12}) and get rid of the integral by 
introducing a function $Y(x)$, such that we have
\begin{equation}
\label{eq14}
g(x) = {Y}'(x),\qquad Y(x) = {\int\limits_{x_{0}} ^{x} {g(}} \xi )\,d\,\xi .
\end{equation}
The $\Phi $--functions can then be expressed in terms of $Y(x)$ and of some 
constants as
\bea
\label{eq15}
 \Phi _{2} (x) &=& {\frac{{1}}{{{Y}'(x)^{2}}}}{\left[ {Y(x) + c_{0}}  
\right]}, \\ \nonumber
 \Phi _{1} (x) &=& {\frac{{1}}{{{Y}'(x)^{3}}}}{\left[ { - 3\,(Y(x) + c_{0} 
){Y}''(x) + (c_{1} + 3)\,{Y}'(x)^{2}} \right]}, \\ \nonumber
 \Phi _{0} (x) &=& - {\frac{{1}}{{{Y}'(x)^{4}}}} [ (Y(x) + c_{0} 
)({Y}'''(x){Y}'(x) - 3\,{Y}''(x)^{2}) \\ \nonumber && + (c_{1} + 3)\,{Y}'(x)^{2}{Y}''(x)
]. 
\eea
The boundary condition (\ref{eq11}) can be rewritten in terms of the function $Y(x)$ (see below). The operators $\hat P$ and $\hat M$ should be applied only to functions $u(x)$ satisfying this boundary condition.

The results of this section can be summed as a theorem

\begin{theorem} \label{t1} The operators
\begin{equation}
\label{eq16}
\begin{array}{l}
 \hat {P} = {\frac{{1}}{{{Y}'(x)^{2}}}}{\left[ {Y(x) + c_{0}}  
\right]}\hat {D}_{x}^{2} + {\frac{{1}}{{{Y}'(x)^{3}}}}{\left[ {(c_{1} + 3){Y}'(x)^{2}} - 3(Y(x) 
+ c_{0} ){Y}''(x)  \right]}\hat {D}_{x}  \\ 
 - {\frac{{1}}{{{Y}'(x)^{4}}}}{\left[ {(Y(x) + c_{0} )({Y}'''(x){Y}'(x) - 
3\,{Y}''(x)^{2}) + (c_{1} + 3)\,{Y}'(x)^{2}{Y}''(x)} \right]}, \\ \\
 \hat {M} = {Y}'\,\hat {D}_{x}^{ - 1} + k_{0}, \\ 
 \end{array}
\end{equation}
satisfy the Heisenberg relation (\ref{eq4}) and both $\hat {P}\,\hat {M}$ and 
$\,\hat {M}\,\hat {P}$ are second order differential operators. The function 
$Y(x)$ satisfies the condition $Y(x_{0} ) = 0$ and is otherwise arbitrary. The 
constants $c_{0} ,\,c_{1} ,\,x_{0} ,\,k_{0} $ are arbitrary. The operators 
$\hat {P}$ and $\hat {M}$ are defined for functions $u(x)$ satisfying the 
boundary condition
\bea
\label{eq17}
&& \biggl [ \frac{1}{Y'(x)^3} \bigl \{ \,(Y(x) + c_{0} 
)\, Y'(x)\, u'(x) + [ \,(2 + c_{1} )\, Y'(x)^{2} \\ \nonumber && \qquad - (Y(x) + c_{0} 
)\,{Y}''(x) ] \,u(x)\, \bigr \} \biggr ]_{x = x_{0}}  = 0.
\eea
\end{theorem}
The inverse of the theorem is also true. All operators $\hat {P},\,\hat 
{M}$ satisfying the above properties are given by (\ref{eq16}).

\section{MONOMIALITY AND BASIS FUNCTIONS}

\subsection{General approach}

The fundamental notion underlying monomiality is the existence of a sequence 
of basis functions $u_{n} (x)$satisfying
\begin{equation}
\label{eq18}
\begin{array}{l}
 \hat {P}\,u_{n} (x) = n\,u_{n - 1} (x), \\ 
 \hat {M}\,u_{n} (x) = u_{n + 1} (x), \\ 
 \end{array}
\end{equation}
these two equations imply
\begin{equation}
\label{eq19}
\begin{array}{l}
 \hat {P}\,\hat {M}\,u_{n} (x) = (n + 1)\,u_{n} (x), \\ 
 {\left[ {\hat {P},\,\hat {M}} \right]}\,\,u_{n} (x) = \hat {1}\,u_{n} (x) 
\\ 
 \end{array}
\end{equation}
The functions $u_n(x)$ are given explicitly as ``monomials'' 
\begin{equation}
\label{eq20}
u_{n} (x) = \hat {M}^{n}\,u_{0} (x),
\end{equation}
and in terms of the operator $\hat {M}$ and some ``seed function'' or "vacuum function" $u_{0} (x)$, 
which is to be defined.
The use of the Heisenberg relation along with eq. (\ref{eq20}) yields
\begin{equation}
\label{eq21}
\hat {P}\,u_{n} (x) = n\,u_{n - 1} (x) + \hat {M}^{n}(\hat {P}\,u_{0} (x)).
\end{equation}
To satisfy the first of eq. (\ref{eq18}) for all $n$ including $n=0$, we must impose
\begin{equation}
\label{eq22}
\hat {P}\,u_{0} (x) = 0.
\end{equation}
The above conditions can be viewed in at least two different ways

\begin{description}
\item[a] As a condition on the operator $\hat {P}$ and thus on the function $Y(x)$
\item[b] As a condition on the seed function\textit{} $u_{0} (x)$.
\end{description}
It is evident that $u_{0} (x)$ plays the role of a physical vacuum.

\subsection{Conditions on the operator$\hat {P}$ for a constant 
seed function}

Let us put $u_{0} = 1$ (or any other non-zero constant). Condition (\ref{eq22}) 
implies $\Phi _{0} (x) = 0$ that is, in view of eq. (\ref{eq16}) 
\begin{equation}
\label{eq23}
(Y(x) + c_{0} )\,({Y}'(x)\,{Y}'''(x) - 3\,{Y}''(x)^{2}) + (c_{1} + 
3){Y}'(x)^{2}{Y}''(x) = 0.
\end{equation}
The function $Y(x)$ is therefore no longer arbitrary, but it is subject to 
eq. (\ref{eq23}). This equation has a three-dimensional Lie point symmetry group, 
generated by the Lie algebra
\begin{equation}
\label{eq24}
\begin{array}{l}
 \hat {X}_{1} = \partial _{x} , \\ 
 \hat {X}_{2} = x\,\partial _{x} , \\ 
 \hat {X}_{3} = (Y + c_{0} )\,\partial _{Y} \\ 
 \end{array}
\end{equation}
which can be used to reduce eq. (\ref{eq23}) to quadratures. However we obtain implicit
solutions of little use in the present context. We therefore use an 
alternative approach for $u_{0} = 1$, namely we return to the equations 
solved in Section 2 and impose $\Phi _{0} (x) = 0$ from the beginning. Eqs. 
(\ref{eq5}, \ref{eq9}) with $\Phi _{0} (x) = 0$ imply
\begin{equation}
\label{eq25}
\begin{array}{l}
 \Phi _{1} (x) = {\Phi} '_{2} (x) + \alpha , \\ 
 g(x) = {\frac{{1 - 2\,\beta} }{{{\Phi} '_{2} (x) - 2\,\alpha} }},\qquad {\Phi 
}'_{2} (x) \ne 2\,\alpha ,\qquad \beta \ne {\frac{{1}}{{2}}} \\ 
 \end{array}
\end{equation}
where $\alpha$, $\beta$ are integration constants. Eq. (\ref{eq10}) implies that 
$\Phi _{2} (x)$satisfies the condition
\begin{equation}
\label{eq26}
(1 - 2\,\beta )\,\Phi _{2} (x)\,{\Phi} ''_{2} (x) + \beta {\left[ {{\Phi 
}'_{2} (x)} \right]}^{2} - \alpha \,(1 + 2\,\beta )\,{\Phi} '_{2} (x) + 
2\,\alpha ^{2} = 0.
\end{equation}
By solving the above equation for $\alpha = 0$ we obtain a simple but 
interesting solution
\begin{equation}
\label{eq27}
\begin{array}{l}
 \Phi _{2} (x) = {\frac{{(x + \gamma )^{1 - q}}}{{A (1 + 
q)}}},\quad \Phi _{1} (x) = {\Phi} '_{2} (x) = {\frac{{1 - q}}{{A (1 + 
q)}}}\,\left( {x + \gamma}  \right)^{ - q},\quad \Phi _{0} (x) 
= 0, \\ 
 q \ne - 1, \qquad \beta = \frac{q}{1+q}. \\ 
 \end{array}
\end{equation}
Eqs. (\ref{eq15}) then imply that 
\begin{equation}
\label{eq28}
\begin{array}{l}
 {Y}'(x) = A\,(x + \gamma )^{q},\qquad c_{0} = {\frac{{A}}{{q + 1}}}(x_{0} + \gamma )^{q + 1}, \\ 
Y(x) = {\frac{{A}}{{q + 1}}}{\left[ 
{(x + \gamma )^{q + 1} - (x_{0} + \gamma )^{q + 1}} \right]} \,, \quad c_{1} = - 
{\frac{{q + 2}}{{q + 1}}}. \\ 
 \end{array}
\end{equation}
The operator $\hat {P}$ reduces to
\begin{equation}
\label{eq29}
\hat {P} = {\frac{{1}}{{A\,(q + 1)}}}\,\hat {D}_{x} (x + \gamma )^{1 - 
q}\hat {D}_{x} 
\end{equation}
and is self-adjoint.

The boundary condition (\ref{eq17}) is satisfied identically for $u(x) =  
const$ once the constants $c_{0,1}$ are those given in eq. (\ref{eq28}).

\subsection{Condition on the seed function $u(x)$ for arbitrary 
$Y(x)$} 

Let us now consider the operator $\hat {P}$ of eq. (\ref{eq16}) with $Y(x)$ (and 
hence $g(x) = {Y}'(x)$) arbitrary. The seed function must satisfy eq. (\ref{eq22}). 
From eq. (\ref{eq5}) we see that $u_{0} (x) = {Y}'(x)$ is a solution of eq. (\ref{eq22}). 
A second linearly independent solution is easily obtained using the 
Wronskian. The general solution of eq. (\ref{eq22}) can be written as
\begin{equation}
\label{eq30}
\begin{array}{l}
 u_{0} (x) = {Y}'(x){\left[ {a{}_{1} + a_{2} (Y(x) + c_{0} )^{ - c_{1} - 2}} 
\right]}\,,\quad c{}_{1} \ne - 2, \\ 
 \\ 
 \end{array}
\end{equation}
or 
\begin{equation}
\label{eq31}
u_{0} (x) = {Y}'(x){\left[ {a{}_{1} + a_{2} \ln (Y(x) + c_{0} )} 
\right]}\,,\quad c{}_{1} = - 2.
\end{equation}
Since we are dealing with linear equations we can consider separately the 
cases $a_{1} = 1$, $a_{2} = 0$ and $a_{1} = 0$, $a_{2} = 1$ when imposing the 
boundary condition (\ref{eq17}).
The boundary condition is satisfied for $ u_{0} (x) = {Y}'(x) (Y(x) + c_{0} )^{ - c_{1} - 2}$ 
for all values of $c_1$, in particular for the case $c_1 = -2$, i.e. $u_0(x)=Y'(x)$. It is not satisfied for the term with a logarithm, so we discard the solution (\ref{eq31}).

\subsection{The eigenvalue problem for the basis functions $u_{n} (x)$}

The functions $u_{n} (x)$ defined by the relations (\ref{eq18}, \ref{eq19}, \ref{eq20}) 
with $u_{0} (x)$ satisfying eq. (\ref{eq22}), will be eigenfunctions of the linear 
operator $\hat {L} = \hat {P}\,\hat {M}$ and the situation can be summed up 
according to the following theorem
\begin{theorem} \label{t2}
The functions $u_{n} (x)$ defined by
\begin{equation}
\label{eq32}
\begin{array}{l}
 u_{n} (x) = \hat {M}^{n}u_{0} (x), \\ 
 \hat {P}\,u_{0} (x) = 0, 
 \end{array}
\end{equation}
satisfy the second order \textit{ODE}
\begin{equation}
\label{eq33}
\begin{array}{l}
 \hat {L}\,u_{n} (x) = (n + 1)\,u_{n} (x), \\ 
 \hat {L} = f_{2} (x)\,\hat {D}_{x}^{2} + f_{1} (x)\,\hat {D}_{x} + f_{0} 
(x), 
 \end{array}
\end{equation}
with
\begin{equation}
\label{eq34}
\begin{array}{l}
 f_{2} (x) = {\frac{{k_{0}} }{{{Y}'(x)^{2}}}}{\left[ {Y(x) + c_{0}}  
\right]}, \\ 
 f_{1} (x) = {\frac{{1}}{{{Y}'(x)^{3}}}}{\left\{ {\,{\left[ {Y(x) + c_{0} + 
k_{0} (c_{1} + 3)} \right]}\,\,{Y}'(x)^{2} - 3\,k_{0} (Y(x) + c_{0} 
)\,{Y}''(x)} \right\}}, \\ 
 f_{0} (x) =- {\frac{{1}}{{{Y}'(x)^{4}}}}\,{\left\{ {\begin{array}{l}
 {k_{0} (Y(x) + c_{0} )\,({Y}'(x)\,{Y}'''(x) - 3\,{Y}''(x)^{2}) +} , \\ 
 { + {\left[ {k_{0} (c_{1} + 3) + Y(x) + c_{0}}  
\right]}\,{Y}'(x)^{2}{Y}''(x) -}  \\ 
 { - (c_{1} + 3)\,{Y}'(x)^{4}} \\ 
 \end{array}}
 \right\}} ,
 \end{array}
\end{equation}
as long as $Y(x)$ satisfies the boundary condition (\ref{eq17}). The operators $\hat M$, $\hat P$ are 
those of Theorem \ref{t1}.  The seed function $u_{0} (x)$ must be chosen as in 
eq. (\ref{eq30}) for $Y(x)$ arbitrary, or $u_0(x)=1$ for $Y(x)$ as in eq. (\ref{eq28}).
\end{theorem}
The seed function $u_{0} (x)$ satisfies eq. (\ref{eq17}), but this does 
not guarantee the same for all $u_{i} (x),\,i \geqslant 1$.

 We will see in 
Section 5 that this imposes a condition on the constants $c_{0,\,1} $.

\subsection{Eigenfunctions as functions of two variables.}

We have constructed the monomials $u_n(x) = \hat M^n u_o$ as functions of one variable $x$.  They also depend in a significant way on the parameter $k_0$. Let us, for the purposes of this paragraph, change notation and write
\bea \label{3.20}
\pi_n(x,y) \equiv u_n(x), \quad y \equiv k_0.
\eea
We then have
\bea \label{3.21}
\pi_n(x,y) = \hat M^n u_0, \qquad \hat M = \dot Y(x) D_x^{-1} + y.
\eea

The functions $\pi_n(x,y)$, as functions of two variables, satisfy a partial differential equation
\bea \label{3.22}
\frac{\partial \pi_n(x,y)}{\partial y} = \hat P \pi_n(x,y).
\eea
Indeed, we have
\bea \nonumber
\frac{\partial \pi_n(x,y)}{\partial y} = n \hat M^{n-1} \frac{\partial \hat M}{\partial y} u_0.
\eea
We have  $\partial \hat M/\partial y = 1$, and eq. (\ref{eq18}) leads to eq. (\ref{3.22}). Eq. (\ref{3.22}) can be written explicitly as a second order partial differential equation, using eq. (\ref{eq16}). In particular, for a constant seed function $u_0=1$ we have $\hat P$ as in eq. (\ref{eq29}) and eq. (\ref{3.22}) reduces to
\bea \label{3.23}
\frac{\partial \pi_n(x,y)}{\partial y} = \frac{1}{A (1 + q)} D_x x^{1-q} D_x \pi_n, \quad q \ne -1,
\eea
where we have put $\gamma=0$.

Eq. (\ref{3.23}) is a linear heat equation with variable conductivity $x^{1-q}$ and the monomials $u_n(x) = \pi_n(x,y) = \hat M^n u_0$ are solutions of this equation.

Let us put
$$ y = A (1+q)t, \qquad 1-q = N, \qquad q \ne -1 \,\, (N \ne 2).$$
Eq.(\ref{3.23}) now is 
\bea \label{3.9b}
\frac{\partial u(x,t)}{\partial t} = D_x x^N D_x u(x,t).
\eea
Its Lie point symmetry algebra can be calculated using standard methods \cite{13b}. For $N=0$ and $N=\frac{4}{3}$ the algebra is six dimensional and the $N=\frac{4}{3}$ case is isomorphic to the $N=0$ one, i.e. to the symmetry algebra of the constant coefficient heat equation. A basis for the symmetry algebra for $N=0$ and $N=\frac{4}{3}$ is
\bea \label{40b}
X_1 &=& \partial_t, \\ \nonumber
X_2 &=& t\partial_t +\frac{1}{2-N} x \partial_x - \frac{1}{2 (2-N)} u \partial_u, \\ \nonumber
X_3 &=& t^2\partial_t +\frac{2}{2-N} t  x \partial_x - \bigl [ \frac{1}{(2-N)^2} x^{2-N} + \frac{1}{2-N} t \bigr ] u \partial_u, \\ \nonumber
X_4 &=& x^{\frac{N}{2}} \partial_x - \frac{N}{4} x^{\frac{N-2}{2}}  u \partial_u, \\ \nonumber
X_5 &=& t x^{\frac{N}{2}} \partial_x - \bigl [ \frac{N}{4} t x^{\frac{N-2}{2}} + \frac{1}{2-N} x^{\frac{2-N}{2}} \bigr ]  u \partial_u, \\ \nonumber
X_6 &=& u \partial_u.
\eea

The fact that the two algebras are isomorphic suggests that the two equations could be transformed into each other by a point transformation (it is a necessary condition for the existence of such a transformation). This is indeed the case here. We put
\bea \label{41b}
y=t, \quad z=3 x^{1/3}, \quad w(z,y) = x^{1/3} u(x,t).
\eea
Then, if $u(x,t)$ satisfies
\bea \label{42b}
\frac{\partial u}{\partial t} = \frac{\partial}{\partial x} x^{4/3} \frac{\partial u}{\partial x},
\eea
$w(z,y)$ will satisfy
\bea \label{43b}
w_y = w_{zz}
\eea
and vice versa.

For $N \ne 0, \frac{4}{3}, 2$ the symmetry algebra of eq, (\ref{3.9b}) is four--dimensional with basis given by $X_1$, $X_2$, $X_3$ and $X_6$ of eq. (\ref{40b}). The symmetry group in this general case is $GL(2,\mathcal R)$ and the equation (\ref{3.9b}) can not be transformed into the usual heat equation (\ref{43b}).

\section{ ORTHOGONALITY PROPERTIES OF THE EIGENFUNCTIONS}

Let us first recall some well known results from the theory of linear 
operators \cite[14]. The adjoint $\hat {L}^{ +}$ of the second linear differential 
operator $\hat {L}$ of eq. (\ref{eq33}) is defined by the relation
\begin{equation}
\label{eq35}
{\int\limits_{a}^{b} {U_{2}} } (x)\,\hat {L}\,\,U_{1} (x)\,dx = 
{\int\limits_{a}^{b} {U_{1}} } (x)\,\hat {L}^{ +} U_{2} (x)\,dx,
\end{equation}
where $U_{2,\,1} (x)$ satisfy the boundary condition, 
\begin{equation}
\label{eq36}
\left( {U{}_{2}(x)\,f_{2} (x)\,{U}'_{1} (x) - U_{1} (x)\,\left( {f_{2} 
(x)\,U_{1} (x)} \right)^{\prime}  + f_{1} (x)\,U{}_{1}(x)\,U_{2} (x)} 
\right)\vert _{a}^{b} = 0,
\end{equation}
and we have 
\begin{equation}
\label{eq37}
\hat {L}^{ +}  = f_{2} (x)\,\hat {D}_{x}^{2} + (2\,{f}'_{2} (x) - f_{1} 
(x))\,\hat {D}_{x}  + f_2'' - {f}'_{1} (x) + f_{0} (x).
\end{equation}
Let us introduce the eigenfunctions of $\hat {L}$ and $ \hat 
{L}^{ +} $
\begin{equation}
\label{eq38}
\begin{array}{l}
 \hat {L}\,u_{n} (x) = \lambda _{n} \,u_{n} (x), \\ 
 \hat {L}^{ +} v_{m} (x) = \lambda _{m} v_{m} (x) .
 \end{array}
\end{equation}
The functions $u_{n} (x),\,v_{m} (x)$ form mutually orthogonal sets
\begin{equation}
\label{eq39}
{\int\limits_{a}^{b} {v_{m}} } (x)\,u_{n} (x)\,dx = 0,\,m \ne n\quad \lambda 
_{m} \ne \lambda _{n} ,
\end{equation}
where $v_{m} (x)$ (in the domain of $\hat {L}^{ +} $) and $u_{n} (x)$ (in 
the domain of $\hat {L}\,$) must satisfy the same boundary conditions (\ref{eq36}) 
(for some $a$ and $b$ ). Now let us request that the eigenfunctions $v_{n} (x)$ be 
proportional to $u_{n} (x)$
\begin{equation}
\label{eq40}
v_{n} (x) = w(x)\,u_{n} (x).
\end{equation}
This implies that $w(x)$ must satisfy the following first order \textit{ODE}
\begin{equation}
\label{eq41}
f_{2} (x)\,{w}'(x) + ({f}'_{2} (x) - f_{1} (x))\,w(x) = 0.
\end{equation}
The boundary condition for $U_{1} (x) = u_{n} (x)$and $U_{2} (x) = v_{m} (x) 
= w(x)\,u_{m} (x)$ reduces to
\begin{equation}
\label{eq42}
{\left[ {f_{2} (x)\,w(x)(u_{m} (x)\,{u}'_{n} (x) - {u}'_{m} (x)\,u_{n} (x))} 
\right]}\,_{a}^{b} = 0.
\end{equation}
If $w(x)$ satisfies (\ref{eq41}), (\ref{eq42}) then the eigenfunctions $u_{n} (x)$ of the 
original operator $\hat {L}$ will be orthogonal with the weight $w(x)$
\begin{equation}
\label{eq43}
{\int\limits_{a}^{b} {w(x)\,u_{m}} } (x)\,u_{n} (x)\,dx = 0,\quad \quad 
\lambda _{m} \ne \lambda _{n}. 
\end{equation}
Applying the above general results to the operator $\hat {L}$ of (\ref{t2}) we 
obtain the following theorem
\begin{theorem} \label{t3}
The eigenfunctions $u_{n} (x)$ of eq. (\ref{eq33}) will satisfy the orthogonality 
relation
\begin{equation}
\label{eq44}
{\int\limits_{a}^{b} {w(x)\,u_{m}} } (x)\,u_{n} (x)\,dx = N_{n\,m} \delta 
_{n\,m} ,
\end{equation}
with
\begin{equation}
\label{eq45}
w(x)\, = {\frac{{(Y(x) + c_{0} )\,^{c_{1} + 2}}}{{{\left| {{Y}'(x)} 
\right|}}}}e^{{\frac{{1}}{{k_{0}} }}\,Y(x)},
\end{equation}
provided they satisfy the boundary conditions
\begin{equation}
\label{eq46}
\left\{ \frac{(Y(x) + c_0 )^{c_1 + 3}}{\left|  Y'(x) 
\right|\,^3} \, e^{\frac{1}{k_0} \,Y(x)} \, \left ( u_m (x)\, u'_n (x) - u'_m (x)\,u_n (x)\right ) 
\right\}_a^b = 0.
\end{equation}
\end{theorem}

We mention that the boundary condition (\ref{eq46}) for orthogonality and (\ref{eq17}) 
for (\ref{t1}, \ref{t2}) to hold must be both satisfied and that $x_{0}$,
$a$, $b$ are apriori independent constants (to be determined below).

\section{EXPLICIT FORMS OF THE EIGENFUNCTIONS}

\subsection{Eigenfunctions for $u_{0} (x) = {Y}'(x)\,{\left[ {Y(x) + c_{0} 
} \right]}\,^{\alpha} $  with $Y(x)$\textbf{arbitrary}}

Here we proceed in the spirit of Section (3.3) and put
\begin{equation}
\label{eq47}
c_{1} = - 2 - \alpha ,\quad u_{0} (x) = {Y}'(x)\,(Y(x) + c_{0} )\,^{\alpha} 
\end{equation}
i.e. we start from the second solution in eq. (\ref{eq30}). The first one in 
(\ref{eq30}) is just the special case of $\alpha = 0$. The condition (\ref{eq22}) is 
satisfied as is the boundary condition (\ref{eq17}) for $u_{0} (x)$. The next 
eigenfunction calculated according to eq. (\ref{eq20}) is
\begin{equation}
\label{eq48}
u_{1} (x) = {Y}'(x)\,{\left[ {{\frac{{(Y(x) + c_{0} )\,^{\alpha + 
1}}}{{\alpha + 1}}} + k_{0} (Y(x) + c_{0} ) - {\frac{{1}}{{\alpha + 
1}}}\,c_{0}^{\alpha + 1}}  \right]}.
\end{equation}
Substituting into the boundary condition (\ref{eq17}) we obtain $c_{0}^{\alpha + 
1} = 0$. It follows that we have
\begin{equation}
\label{eq49}
c_{0}^{} = 0,\quad \alpha > - 1.
\end{equation}
Using eq. (\ref{eq20}) we can easily generate $u_{2} (x),\,u_{3} (x),\,...$. 
Inspired by the form of these functions we make the Ansatz
\begin{equation}
\label{eq50}
u_{n} (x) = {Y}'(x)\,Y(x)^{\alpha} g_{n} ( - {\frac{{Y(x)}}{{k_{0}} }}),
\end{equation}
and substitute it into the eigenvalue equation (\ref{eq33}). We find that the function 
$g_{n} $ satisfies the generalised, or associated \cite{15,16} Laguerre equation
\begin{equation}
\label{eq51}
( - {\frac{{Y(x)}}{{k_{0}} }})\,{g}''_{n} + {\left[ {\alpha + 1 - ( - 
{\frac{{Y(x)}}{{k_{0}} }})} \right]}\,{g}'_{n} (x) + n\,g_{n} (x) = 0
\end{equation}
Thus eq. (\ref{eq33}) is solved in terms of generalised Laguerre polynomials 
$L_{n}^{(\alpha )} (z)$. Substituting (\ref{eq50}) into the boundary condition 
(\ref{eq17}), we see that (\ref{eq17}) is satisfied for all values of $n$. The boundary 
condition (\ref{eq46}) for orthogonality is satisfied for all $n$ and $m$ if we choose 
\begin{equation}
\label{eq52}
a = 0,\,\;b = \infty ,\quad {\left[ {{\frac{{1}}{{k_{0}} }}Y(x)} \right]}_{x\, 
\to \infty}  \to - \infty 
\end{equation}
(we have put, with no loss of generality, $x_{0} = 0$).
Eq. (\ref{eq32}) gives us the explicit form of the eigenfunctions for all $n.$

Let us sum up the previous results as the following theorem
\begin{theorem} \label{t4}
The eigenvalue problem 
\begin{equation} \label{5.7}
\begin{array}{l}
 \frac{k_0}{Y'(x)^2}Y(x) u''_n (x) + 
\frac{1}{Y'(x)^3} \left \{ \left[ Y(x) + k_0 (1 - \alpha ) 
\right] Y'(x)^2 - 3 k_0 Y (x) Y''(x) 
\right\} u'_n(x)  \\ 
 - \frac{1}{Y'(x)^4} \biggl \{ k_0 Y(x)\,\left( 
Y'(x)\,Y'''(x) - 3\, Y''(x)^2 \right) + \left[ k_0 (1 - \alpha 
) + Y(x) \right] \,Y'(x)^2 Y''(x) \\ - (1 - \alpha )\, Y'(x)^4 
\biggr \} \,u_n =  
  (n + 1)\,u_n (x),
 \end{array}
\end{equation}
with the boundary condition
\begin{equation}
\label{eq53}
{\left\{ {{\frac{{1}}{{{Y}'(x)^{3}}}}\,{\left[ {Y(x)\,{Y}'(x)\,{u}'(x) - 
(\alpha \,{Y}'(x)^{2} + Y(x)\,{Y}''(x))\,u(x)} \right]}} \right\}}\vert _{x 
= 0} =0,
\end{equation}
is solved by the monomials
\begin{equation}
\label{eq54}
 u_{n} (x) = \hat {M}^{n}u_{0} (x), \quad
 u_{0} (x) = {Y}'(x)\,Y\left( {x} \right)^{\alpha} , \quad 
 \hat {M} = {Y}'(x)\,\hat {D}_{x}^{ - 1} + k_{0} .
\end{equation}
Explicitly the solutions are
\bea
\label{eq55}
 u_{n} (x) &=& {Y}'(x)\,Y(x)^{\alpha} {\sum\limits_{j = 0}^{n} 
{{\frac{{1}}{{j!}}}}} \left( {{\begin{array}{*{20}c}
 {n + \alpha}  \hfill \\
 {n - j} \hfill \\
\end{array}} } \right)\,Y(x)^{j}k_{0}^{n - j} \\ \nonumber
 &=& {Y}'(x)\,Y(x)^{\alpha} L_{n}^{(\alpha )} \left( { - {\frac{{Y(x)}}{{k_{0} 
}}}} \right),\quad \alpha > - 1 ,
\eea
where $L_{n}^{(\alpha )} \left( {z} \right)$is a generalized Laguerre polynomial. The 
function $Y(x)$ and the constants $k_{0}$, $\alpha$ satisfy
\begin{equation}
\label{eq56}
\lim _{x \to \infty}  {\frac{{Y(x)}}{{k_{0}} }} = - \infty ,\quad \alpha > - 
1.
\end{equation}
The orthogonality relation is
\begin{equation}
\label{eq57}
{\int\limits_{0}^{\infty}  {{\frac{{Y(x)^{ - \alpha}}}{{{\left| {{Y}'(x)} 
\right|}}}}}} e^{{\frac{{1}}{{k_{0}} }}Y(x)}u_{n} (x)\,u_{m} (x)\,dx = ( - 
k_{0} )^{\alpha} \delta _{m,n} 
\end{equation}
\end{theorem}

\subsection{Eigenfunctions for $u_{0} (x) = 1,\,{Y}'(x) = A\,\left( {x + 
\gamma}  \right)\,^{q}$}

\par Let us now take ${Y}(x)$, $c_0$ and $c_1$ as in eq. (\ref{eq28}).  The eigenvalue problem (\ref{eq33}) reduces to 
\begin{equation}
\label{eq58}
\begin{array}{l}
 {\frac{{k_{0}} }{{A\,(q + 1)}}}\,(x + \gamma )^{1 - q}\hat {D}_{x}^{2} 
u_{n} (x) + {\frac{{1}}{{A\,\left( {q - 1} \right)}}}{\left\{ {A\,\left( {x 
+ \gamma}  \right) - k_{0} (q - 1) (x + \gamma)^{ - q}} \right\}}\hat {D}_{x} u_{n} (x) = 
\\ 
 = n\,u_{n} (x),  
 \end{array}
\end{equation}
the boundary condition (\ref{eq17}) reduces to
\begin{equation}
\label{eq59}
(x + \gamma )^{1 - q}{u}'(x)\vert _{x = x_{0}}  = 0.
\end{equation}
Condition (\ref{eq59}) is satisfied for $u_{0} (x) = 1$.

We have
\begin{equation}
\label{eq60}
u_{1} (x) = \hat {M}\,u_{0} (x) = A\,(x + \gamma )^{q}(x - x_{0} ) + k_{0} 
\end{equation}
and eq. (\ref{eq59}) for $u_{1} (x)$implies 
\begin{equation}
\label{eq61}
\gamma = - x_{0} ,\quad c_{0} = 0,
\end{equation}
so that we find 
\begin{equation}
\label{eq62}
 \hat {P} = {\frac{{1}}{{A\,(q + 1)}}}\,\hat {D}_{x} (x - x_{0} )^{1 - 
q}\hat {D}_{x} , \quad
 \hat {M} = A\left( {x - x_{0}}  \right)^{q}\hat {D}_{x}^{ - 1} + k_{0} .
\end{equation}
We can easily implement eq. (\ref{eq20}) and get
\begin{equation}
\label{eq63}
u_{n} (x) = {\sum\limits_{j = 0}^{n} {\left( {{\begin{array}{*{20}c}
 {n} \hfill \\
 {j} \hfill \\
\end{array}} } \right)\,k_{0}^{n - j}} } {\frac{{A^{j}(x - x_{0} )^{j(q + 
1)}}}{{(q + 2)\,(2\,q + 3)...((j - 1)\,q + j)}}}
\end{equation}

Substituting $u_{n}$ into the boundary condition (\ref{eq59}) with $\gamma = - 
x_{0}$ we obtain the condition 
\begin{equation}
\label{eq64}
q > - 1.
\end{equation}
Eq (\ref{eq63}) allows us to relate  $u_{n} (x)$ to the generalised Laguerre 
polynomials according to the identity
\begin{equation}
\label{eq65}
\begin{array}{l}
 u_{n} (x) = {\frac{{\Gamma ({\frac{{1}}{{1 + q}}})\,n!}}{{\Gamma (n + 
{\frac{{1}}{{1 + q}}})}}}\,k_{0}^{n} \,L_{n}^{(\alpha )} (z) \\ 
 \alpha = - {\frac{{q}}{{q + 1}}},\quad z = - {\frac{{A}}{{k_{0} \,(q + 
1)}}}x^{q + 1} , 
 \end{array}
\end{equation}
where we have set $x_{0} = 0$, with no loss of generality.

The substitution (\ref{eq65}) reduces the eigenvalue problem (\ref{eq58}) for $\gamma = 
- x_{0} = 0$ to the generalized Laguerre equation for $L_{n}^{(\alpha )} (z)$.
The orthogonality relation (\ref{eq44}) in this case specializes to
\bea
\label{eq66}
&& {\int\limits_{0}^{\infty}  {\left( {{\frac{{A}}{{q + 1}}}} \right)} 
}^{{\frac{{q}}{{q + 1}}}}{\frac{{1}}{{A}}}e^{{\frac{{A\,x^{q + 1}}}{{k_{0} 
(q + 1)}}}}u_{n} (x)\,u_{m} (x) \\ \nonumber && \qquad
 = {\frac{{\left( {\Gamma ({\frac{{1}}{{1 + q}}})} \right)^{2}\,n!}}{{\Gamma 
(n + {\frac{{1}}{{1 + q}}})}}}\left( {{\frac{{A}}{{q + 1}}}} 
\right)^{{\frac{{2\,q}}{{q + 1}}}}{\frac{{1}}{{A^{2}}}}( - k_{0} 
)^{{\frac{{1}}{{q + 1}}}}\delta _{n,\,m} .
\eea
In conclusion we state the following theorem.
\begin{theorem} \label{t5}
The eigenvalue problem 
\begin{equation}
\label{eq67}
\begin{array}{l}
 {u}''_{n} (x) + {\left[ {{\frac{{A}}{{k_{0}} }}x^{q} + (1 - q)\,x^{ - 1}} 
\right]}\,{u}'_{n} (x) - {\frac{{A\,(q + 1)}}{{k_{0}} }}n\,x^{q - 1}u_{n} 
(x) = 0, \\ 
 x^{1 - q}{u}'_{n} (x)\vert _{x = 0} ,\quad q > - 1,
 \end{array}
\end{equation}
is solved by the \textit{monomials} 
\begin{equation}
\label{eq68}
\begin{array}{l}
 u_{n} (x) = \hat {M}^{n}1, \\ 
 \hat {M} = A\,x^{q}\hat {D}_{x}^{ - 1} + k_{0} .
 \end{array}
\end{equation}
The eigenfunctions $u_{n} (x)$ are expressed in terms of generalised Laguerre 
polynomials by eq. (\ref{eq65}). Their normalization is given in eq. (\ref{eq66}).
\end{theorem}

As a further remark let us note that the polynomials we have derived cannot 
be framed within the context of the Sheffer type family, the operator $\hat P$ is 
indeed a function either of $x$ and of the ordinary derivative $\hat {D}_{x} $ 
(see ref. \cite{7} for further comments). More  generally they belong to the 
Laguerre family which provides polynomial forms whose generating function 
cannot be expressed in terms of exponential functions.

\section{APPLICATIONS}

\subsection{Applications to the solution of integro-differential equations}

\par An important aspect of umbral calculus is the {\it umbral correspondence} \cite{1,2,3,4}: the correspondence between results obtained in different realizations of the Heisenberg algebra.

Let us first consider a simple example, namely the initial value problem for a first order partial differential equation
\bea \label{6.1}
\frac{\partial y}{\partial \tau} = ( \frac{\partial}{\partial x} + x ) y, \quad y(0,x) = 1.
\eea
It can be solved by the method of characteristics, or in a more formal way by putting
\bea \label{6.2}
y(\tau,x) = e^{\tau (x + \partial_x)} 1.
\eea
Using the Baker--Campbell--Hausdorff formula\cite{20a}, which in this case reduces to 
\bea \label{6.3}
e^{\tau (x + \partial_x)} = e^{\tau x} e^{\tau \partial_x} e^{-\frac{1}{2} [ \tau x, \tau \partial_x ]},
\eea
we obtain
\bea \label{6.4}
y(\tau,x) = e^{\tau^2/2}e^{\tau x} = e^{\tau^2/2} \sum_{k=0}^{\infty} \frac{1}{k!} \tau^k x^k.
\eea
Now let us replace the PDE (\ref{6.1}) by an operator equation
\bea \label{6.5}
y_{\tau}(\tau,\hat M) = ( \hat P + \hat M ) y(\tau, \hat M),
\eea
where $\hat P$ and $\hat M$ are the operators (\ref{eq2}) studied in the previous sections. In eq. (\ref{6.5}) $y(\tau, \hat M)$ is an operator function. We will apply both sides of eq. (\ref{6.5}) to the seed function $u_0$ and this will turn eq. (\ref{6.5}) into an integro--differential equation for a function $f(\tau,x)$.

A formal solution of the operator equation (\ref{6.5}) is obtained from the expansion (\ref{6.4}) simply by inserting the operator $\hat M$ instead of $x$. We apply this operator $y(\tau, \hat M)$ to the seed function $u_0(x)$ and obtain $f(\tau,x) \equiv y(\tau, \hat M) u_0(x)$. i.e. 
\bea \label{6.6}
f(\tau, x) = e^{\tau^2/2} \sum_{k=0}^{\infty} \frac{1}{k!} \tau^k u_k(x),
\eea
where $u_k(x)$ are the basis functions of Section 5, ultimately expressed in terms of generalized Laguerre polynomials.

The function $f(\tau,x)$ is at least a formal solution of the equation
\bea \label{6.7}
y_{\tau}(\tau,\hat M) u_0(x) = (\hat P + \hat M) y(\tau, \hat M) u_0(x),
\eea in our case the integro--differential equation
\bea \label{6.8}
f_{\tau}(\tau,x) &=& \Phi_2 D_x^2 f(\tau,x) + \Phi_1 D_x f(\tau,x) + (\Phi_0 + k_0) f(\tau,x) \\ \nonumber && + \dot Y \int_0^x f(\tau,\tilde x) d \tilde x.
\eea
Here $\Phi_2$, $\Phi_1$ and $\Phi_0$ and $Y$ are the functions determined in Section 2. The formal solution (\ref{6.6}) is a real solution of eq. (\ref{6.8}) if the series converges.

Let us sum up this example. We start from the linear partial diffeential equations (\ref{6.1}) for which we know the solution (\ref{6.4}) of a given initial value problem. We associate an operator equation (\ref{6.5}) and an integro--differential equation (\ref{6.8}) to (\ref{6.1}) via the umbral correspondence. A solution of an initial value problem for (\ref{6.8}) is obtained from (\ref{6.4}) by the umbral correspondence: we replace the powers $x^n$ by the basis functions $u_n(x)$ for the appropriate operators $\hat P$ and $\hat M$ of Sections 3, 4 and 5. The solution (\ref{6.6}) corresponds to the initial value 
\bea \label{6.9}
f(0,x) = u_0(x)
\eea
for eq. (\ref{6.8}).

More generally, let us consider an evolution equation of the form
\bea \label{6.10}
y_{\tau}(\tau,x) = F(\partial_x,x) y(\tau,x), \qquad y(0,x) = f(x),
\eea
where $F(\partial_x,x)$ is a polynomial in $\partial_x$ with coefficients that are power series (or polynomials) in $x$. A solution of the initial value problem (\ref{6.10}) is given by 
\bea \label{6.11}
y(\tau,x) = e^{\tau F(\partial_x,x)} f(x),
\eea
where (\ref{6.11}) can be written (at least in principle) as a power series by applying the Baker--Campbell--Hausdorff formulas in eq. (\ref{6.11}):
\bea \label{6.12}
y(\tau,x) = \sum_{j=0}^{\infty} c_j(\tau) x^j
\eea
(or the power series (\ref{6.12}) can be obtained in any other way).

To the function (\ref{6.12}) we associate another function, namely
\bea \label{6.12a}
f(\tau,x) = y(\tau, \hat M) u_0 = \sum_{j=0}^{\infty} c_j(\tau) u_j(x).
\eea
The function (\ref{6.12a}) will be a solution of the integra--differential equation
\bea \label{6.13}
y_{\tau}(\tau, \hat M) u_0 = F(\hat P, \hat M) y(\tau, \hat M) u_0,
\eea
i.e
\bea \label{6.14}
f_{\tau}(\tau,x) = F(\hat P, \hat M) f(\tau, x),
\eea
with initial condition
\bea \label{91}
f(0,x) = \sum_{j=0}^{\infty} c_j(0) u_j(x).
\eea

In other words, if we know how to solve the PDE (\ref{6.10}), we know how to solve the integro--differential equation (\ref{6.14}).

\subsection{Description of flattened optical beams}

\par In this subsection we will emphasize the extreme usefulness of the above family 
of orthogonal polynomials in applications to optics.

We define the orthogonal function
\bea \label{6.26}
\Phi_n(x,y) = \alpha_n e^{-\frac{A}{2 y (q+1)} x^{q+1}} \pi_n^{(q)}(x,y),
\eea
with $\alpha _{n}$ a suitable normalization constant such that
\begin{equation}
\label{eq93}
{\int\limits_{0}^{\infty}  {\Phi {}_{n}}} (x,y)\,\Phi _{m} (x,y)\,dx = 
\delta _{m,n} .
\end{equation}
The functions $\pi_n^{(q)}(x,y)$ are defined similarly as in Section 3.5, i.e. we put
\bea \label{93}
\pi_n^{(q)}(x,y)= (Ax^qD_x^{-1} + y)^n 1, \quad y = k_0.
\eea
An idea of the shape of the above family of orthogonal functions is given in 
Fig. 1 where we have plotted the first three ``modes'' for $q=3$. 
\begin{figure}
\centering
\includegraphics[width=6.78cm,height=4.92cm]{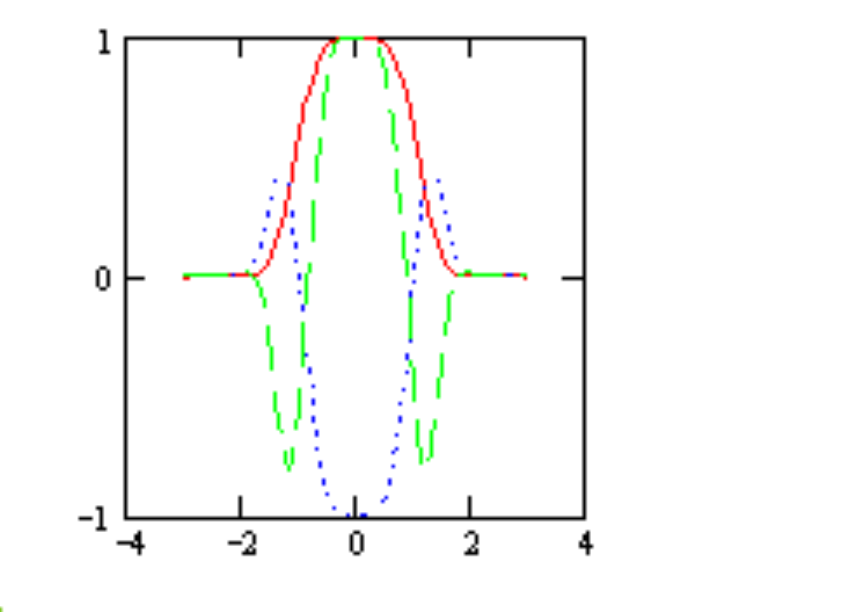}
\caption{The first three flattened $q=3$ beam modes normalised to unity ($n=0$ continuous, 
$n=1$ dot, $n=2$ dash). We take $A=1$ and $y=-1$.} \label{fig2}
\end{figure}
The shape is that of the so called flattened optical beams. These beams are different from the usual Gaussian beams since they have flat transverse distribution as shown in Fig. 2 and Fig. 3.
 \begin{figure}
\centering
\includegraphics[width=6.78cm,height=4.00cm]{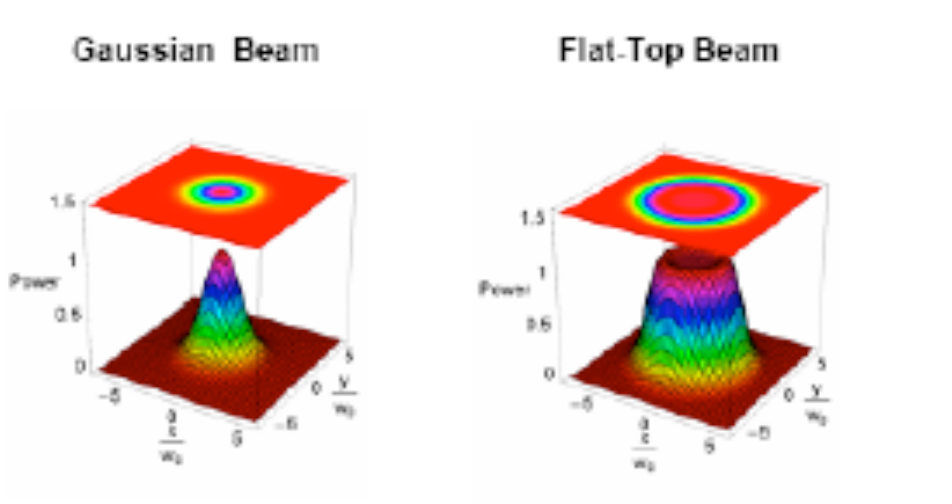}
\caption{Transverse distribution of a Gaussian and a Flattened beam, the difference in intensities  ( denotes the beam waist, with L being the length of the cavity and k the wave vector of the optical field )} \label{fig2a}
\end{figure}
The advantage offered by flattened beams with respect to ordinary Gaussians is that they provide larger suppression of thermal noise on the mirror surface because of a better average on the surface fluctuations, they are therefore particularly useful for high power lasers.
 \begin{figure}
\centering
\includegraphics[width=6.78cm,height=4.00cm]{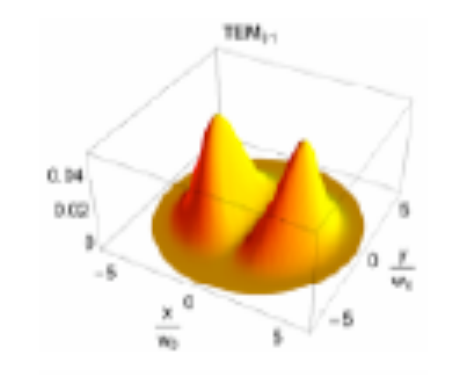}
\caption{Transverse distribution of a higher order flattened mode } \label{fig2b}
\end{figure}
It is evident that if we take a transverse section of the flat top distribution in Figs. 2, 3 we get the distribution similar to that given in Fig. 1 for $n=0$ and $n=1$ respectively \cite{20} used in optics to treat a 
light beam whose cross section has an intensity as uniform as possible. 

A typical example of flattened beam is a super--gaussian
\begin{equation}
\label{eq94}
f(x) \propto e^{ - x^{p}}
\end{equation}
whose profile becomes more and more flat as $p$ increases.

It is evident that the function $\Phi _{n} (x,y)$ is a super--gaussian for 
$n=0$ and that for $n>0$ we have higher order super--gaussian flattened modes.

The distribution of a higher order modes can be obtained from the square 
moduli of the functions (\ref{6.26}) and are shown in Fig. 4 for the cases $n=4$ and $n=30$. 
\begin{figure}
\centering
\includegraphics[width=4.31cm,height=2.92cm]{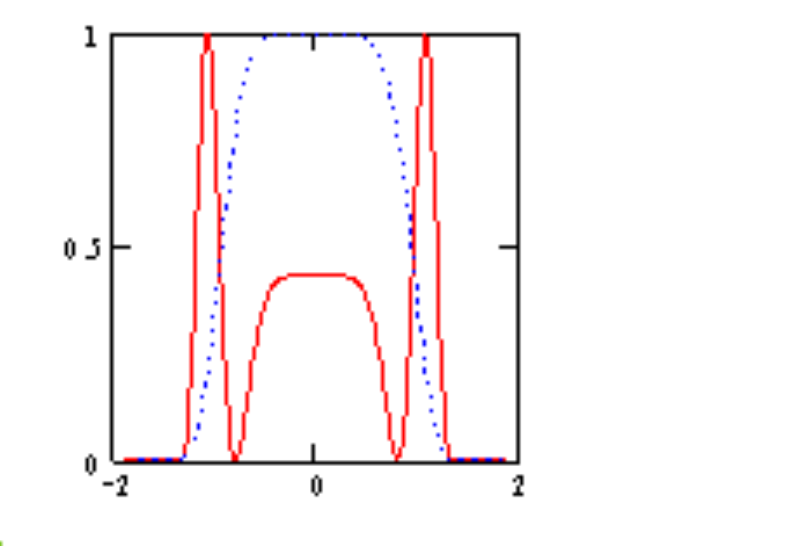}
\includegraphics[width=4.33cm,height=2.88cm]{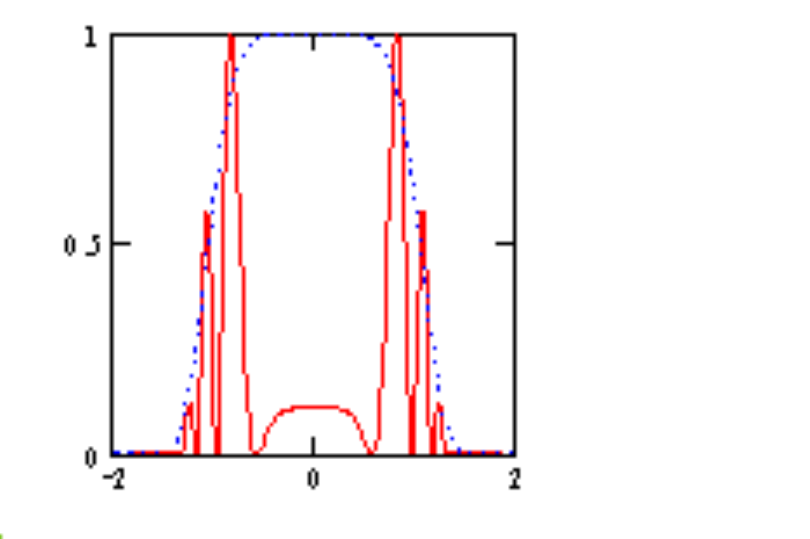}
\caption{Flattened beam mode distribution 
$q=3$ (normalized to unity) and comparison with 
the fundamental. a) $n=4$ ,b) $n=30$} \label{fig3}
\end{figure}

The advantage offered by this family of orthogonal functions is two-fold

\begin{description}
\item[a] They provide the natural set for the expansion of flattened beams
\item[b] They offer the possibility of treating the higher order modes and not only 
the fundamental one.
\end{description}
The study of the propagation of the above family of flattened beams can be 
performed using the set of functions introduced in this article. We postpone this to a   forthcoming investigation.

It is worth pointing out that we have
\begin{equation}
\label{eq95}
\Phi _{n} (x,y) = \alpha _{n} ( - y)^{n}e^{ - 
{\frac{{z}}{{2}}}}L_{n}^{(\alpha )} (z)
\end{equation}
and we can hence discuss the relevant evolution using $z$ as the independent transverse 
coordinate.

It is also worth stressing that the operators  $\hat P, \hat M$ can be used to form other Lie algebras, different from the Heisenberg one, similarly as $x$ and $p=-i/dx$ are used to form e.g. su(1,1). In turn these are well suited  to describe the optical cavity elements and filters 
exploited to flatten the beam distribution, as it will be shown in 
a forthcoming investigation.

\subsection{Applications in Astrophysics}

The Sunyaev--Zeldovich effect \cite{22,23,23b} is the distorsion of the cosmic microwave background radiation spectrum by the inverse Compton scattering of high energy electrons. When describing this effect the authors \cite{22,23} obtained the diffusion equation (the Sunyaev--Zeldovich equation):
\bea \label{95}
\frac{\partial w}{\partial \tau} = \frac{1}{\xi^2} \frac{\partial}{\partial \xi} \xi^4 \frac{\partial w}{\partial \xi}.
\eea

Eq. (\ref{95}) has a symmetry algebra with basis:
\bea \label{96}
X_1 &=& \partial_{\tau} - \frac{9}{4} w \partial_w, \\ \nonumber
X_2 &=& \tau \partial_{\tau} + \frac{1}{2} \xi \log \xi \partial_{\xi} - \frac{1}{4} \bigl [ 9 \tau + 3 \log \xi + 1 \bigr ]  w \partial_w, \\ \nonumber
X_3 &=& \tau^2 \partial_{\tau} + \tau \xi \log \xi \partial_{\xi} - \frac{1}{4} \bigl [ 9 \tau^2 + 2 \tau  + 6 \tau  \log \xi + \log^2 \xi \bigr ]  w \partial_w, \\ \nonumber
X_4 &=& \xi \partial_{\xi} - \frac{3}{2} w \partial_w, \\ \nonumber
X_5 &=& \tau \xi \partial_{\xi} - \frac{1}{2} \bigl [ 3 \tau + \log \xi \bigr ]  w \partial_w, \\ \nonumber
X_6 &=& w \partial_w.
\eea

This Lie algebra is isomorphic to the algebra given in eq. (\ref{40b}), implying that the Sunyaev--Zeldovich equation (\ref{95}) might be equivalent to the heat equation. Indeed it is  and the equivalence is realized by the transformation 
\bea \label{97}
t = \tau, \quad x = \log \xi, \quad u(x,t) = \xi^{3/2} e^{\frac{9}{4} \tau} w(\xi,\tau),
\eea
(see also \cite{nc})
If $w(\xi,\tau)$ satisfies eq. (\ref{95}) then $u(x,t)$ satisfies the heat equation $u_t=u_{xx}$. Hence the Sunyaev--Zeldovich equation can be solved in terms of the functions (\ref{eq63}) constructed in this article (for $q=1$ and $k_0= 2 A t$).
\section{CONCLUSIONS}

One way of summing up the results of the present article is the following. 
We have constructed the most general operators $\hat P, \hat M$ of the form (\ref{eq2}) that 
satisfy the Heisenberg relation (\ref{eq4}) and used them to construct the 
monomials $u_{n} (x) = \hat {M}^{n}u_{0} (x)$. They are eigenfunctions of 
the linear operators $L=\hat P \hat M$ corresponding to positive integer eigenvalues $\lambda 
_{n} = n + 1$ .

The second order operator $L $ given in eq. (\ref{eq33}) is thus factored into a 
product of 2 operators $\hat P $  and $\hat M$. The usual factorization of a differential 
operator is into two lower order differential operators \cite{17,18}. Ours 
is highly non standard: a differential operator $\hat P$ times an integral one $\hat M$.

The functions $u_{n} (x)$ are expressed in terms of generalized Laguerre 
polynomials $L_{n}^{(\alpha )} (z)$, or equivalently , the confluent 
hypergeometric function ${}_{2}F_{0} (a,b;z)$. This includes the Hermite 
polynomials for $\alpha = {\frac{{1}}{{2}}}$, or $\alpha = - 
{\frac{{1}}{{2}}}$ 
but none of the other classical orthogonal polynomials, related to the 
hypergeometric function (rather then the confluent one).

There is a good reason for this. We have imposed the form (\ref{eq2}) on the 
operators $\hat P, \hat M$ and the monomiality condition then leads to an eigenvalue problem 
in which the order of the polynomials $n$ enters in the eigenvalues only and 
enters linearly. For all other classical orthogonal polynomials the 
dependence on $n$ is more general.

We are looking into possible generalizations in order to obtain monomial 
realizations of other classes of orthogonal polynomials. This can e. g. be 
done by imposing all the properties of monomiality, but allowing $\hat P, \hat M$ to depend 
on $n$,
\begin{equation}
\label{eq100}
 \hat {P}_{n} u_{n} (x) = n\,u_{n - 1} (x), \qquad 
 \hat {M}_{n} u_{n} (x) = u_{n + 1} (x).
\end{equation}
The Heisenberg relation should in this case be modified to
\begin{equation}
\label{eq101}
(\hat {P}_{n + 1} \hat {M}_{n} - \hat {M}_{n - 1} \hat P_{n} )\,u_{n} (x) = u_{n} 
(x)
\end{equation}
As also stressed in the previous section the possibility of introducing an 
extra variable makes it possible to consider our polynomials (or functions) as 
depending on two variables (and also on two parameters $n$, $\alpha$).

A further topic that is being pursued is that of applications, as outlined 
in Section 6. A very special case of eq. (\ref{6.14}) was studied in Ref. \cite{8}. We 
intend to study eq. (\ref{6.14}) in general with $\hat P, \hat M$ as specified in this article, for 
a wide choice of the functions $F(\hat P, \hat M)$.

Regarding applications we have stressed that the form of the introduced 
orthogonal polynomials is ideally suited for the study of optical 
flattened beams.

\section*{Acknowledgments}
 D.L thanks the Centro Ricerche ENEA FRASCATI  for its
hospitality during the time this research was realized. DL was partially supported by PRIN Project ÔMetodi geometrici nella teoria delle onde non lineari ed applicazioni-2006Õ of the Italian Minister
for Education and Scientific Research.

The research of P. W. was partially supported by NSERC of Canada. He thanks 
the Centro Ricerche ENEA FRASCATI and the Universit\`{a} ROMA TRE for their 
hospitality and financial support during the time this research was realized.

\end{document}